\input harvmac

\Title{RU-95-90}{\vbox{\centerline{SUSY Breaking, Cosmology, Vacuum}
\centerline{Selection and the Cosmological}\centerline{Constant in 
String Theory }
}}
\bigskip
\centerline{\it 
T. Banks \footnote{*}
{\rm Supported in part by the Department of Energy under grant No.
DE-FG05-90ER40559  .}\footnote{$\dagger$}{\rm Talk Given at the Santa Barbara 
Workshop on Supersymmetry, December 7-9, 1995.} }
\smallskip
\centerline{Department of Physics and Astronomy}
\centerline{Rutgers University}
\centerline{Piscataway, NJ 08855-0849}
\noindent
\bigskip
\baselineskip 18pt
\noindent
This is a summary of a loosely connected collection of results
on supersymmetry (SUSY) breaking and cosmology in the context of superstring
theory.  The requirement of a satisfactory inflationary cosmology puts
strong constraints on the superstring vacuum state.  Some of these
are phenomenological, {\it i.e.} certain things must be true if the
theory is to reproduce observations of the Cosmic Microwave Background.
By far the strongest constraints come from the simple requirement that
the theory have a reasonable probability of producing a large universe.
One can gain a tentative understanding of why there are no more than
four large spacetime dimensions, and of why supersymmetry must be
broken, based on this requirement alone.  The problem of dilaton
domination of the energy density of the universe is discussed and a
plausible resolution due to Lyth and Stewart mentioned. 
Some remarks are made about the pattern of low energy SUSY 
breaking.  In particular, we point out several indications that 
low energy, as opposed to hidden sector SUSY breaking, will pose
cosmological problems in string theory.  If it {\it is} the correct mechanism
then the theory will contain very light particles which will give rise to
coherent forces of millimeter to centimeter range.
Finally, some
speculations based on the holographic principle are advanced which
enable one to estimate the cosmological constant.  If SUSY breaking
comes from a hidden sector the calculation gives a result near the
present observational bound.  For low energy SUSY breaking the result is
too small to be of any observational significance.

\Date{December 1995}

\newsec{\bf Introduction}

In this talk I want to summarize a loosely connected collection of
results that I have obtained in collaboration with M.Dine, M.Berkooz and
others \ref\papers{T Banks, D.Kaplan, A.Nelson, {\it Phys.Rev.}{\bf
D49},(1994), 779,hep-th/9308292;\break
T.Banks, M.Dine, {\it Phys.Rev.}{\bf
D50},(1994),7454,hep-th/9406132; {\it Quantum Moduli\break
Spaces of N = 1 Vacua}, hep-th/9508071, to be published in Phys. Rev. D;
T.Banks, M.Berkooz, P.J.Steinhardt, {\it Phys.Rev.} {\bf D52}, 
(1995),705,hepth/9501053;
T.Banks, M.Berkooz, G.Moore, S.Shenker, P.J.Steinhardt, {\it Phys.
Rev.}{\bf D52},(1995),3548,hep-th/9503114;
T.Banks, L.Susskind, {\it The Number of States of Two Dimensional
Critical String Theory}, RU-95-88, hep-th/9511193. 
}, over the last several years.  The general theme of
these investigations was that the effective field theory of
superstrings (EFTS), informed by cosmological considerations, can give us the
beginnings of an understanding of how the
the Universe chooses the correct vacuum state of superstring theory.
If they succeed in doing nothing else, I hope that these arguments will
convince everyone that the correct context of this question is
cosmological.  The problem is very different from that of solving for
the ground state of a Lorentz invariant field theory that does not
include gravity.

Our work was based on two fundamental assumptions.  The primary one is
the assumption that the EFTS is weakly coupled at a scale just below the
string scale\foot{Another hidden assumption is that the string scale is
indeed the fundamental short distance scale in string theory, and that
other scales like the Planck scale, or the Shenker scale\ref\steve{
S.H.Shenker, {\it Another Length Scale in String Theory?}, RU-95-53,
hep-th/9509132.}
have something do do with the number of degrees of freedom per string
scale.}.  We have several pieces of evidence for this.  The first is the
observed unification of couplings at a value ${g^2 \over 4\pi} \sim
{1\over 25}$.  The second is the striking(?) resemblance of the
perturbative spectra of some string vacua to the spectrum of elementary
particles in the world.  However, I consider the hierarchy problem, the
existence of a wide range of dynamical scales in nature, to be the most
compelling argument in favor of weak coupling.  At present we have no
other way of understanding this fundamental fact than by attributing
it to the properties of marginally relevant gauge couplings, which are
weak at high energies.

The fact that we are in a weakly coupled region leads directly to our
second assumption, namely that we cannot expect cancellation between
different exponentially small effects in the coupling.  Since some of
the most popular models of string phenomenology are based on just such a
cancellation \ref\race{L.Dixon, {\it Supersymmetry Breaking in String
Theory}, Invited talk given at 15th APS Div. of Particles and Fields
Meeting, Houston, TX, Jan. 3-6, 1990;
 V. Krasnikov,
{\it Phys.\ Lett.} {\bf 193B} (1987) 37;
; J.A. Casas, Z. Lalak, C. Munoz and G.G. Ross,
{\it Nucl.\ Phys.} {\bf B347} (1990) 243;
T. Taylor, {\it Phys.\ Lett.} {\bf B252} (1990) 59.}
I should give a definitive argument ruling
these models out.  I will not.  Instead I will characterize my second
assumption as {\it religious}.  I cannot give strong arguments to
justify it, but I believe in it strongly enough that I will let it guide
my future actions.  

Within the context of these assumptions, there is only one known way to
solve the Dine-Seiberg problem\ref\ds{M.Dine, N.Seiberg, {\it Phys.
Lett.}{\bf 162B},(1985),299.} of the runaway dilaton.  The
constraints of analyticity, and the perturbative part of S-duality,
guarantee that the superpotential of the dilaton is well approximated
by a pure exponential in the weak coupling region, and that the dominant
exponential comes from low energy field theory effects \ref\bd{T.Banks,
M.Dine, {\it op. cit.}
} (if any).
The dilaton can nonetheless be stabilized in the region of weak bare
field theory coupling, if its Kahler potential is not well approximated
by the tree level formula.  This is plausible, because Shenker \steve argued
that generic objects in string theory have nonperturbative corrections
of order $e^{-{c\over g}}$.  The results of \bd show that there are no
such corrections to the superpotential, but the argument does not apply
to the Kahler potential.  {\it This means that the perturbation series
for the Kahler potential may break down even for values of the coupling
for which the low energy effective field theory is weakly coupled.}
In \bd a matrix model argument was given suggesting that $c\sim 1$.
This is supported by an explicit calculation of such a nonperturbative
effect by Polchinski\ref\Dinst{J. Polchinski,{\it Phys. Rev.}{\bf
D50},(1994),6041. }. 

A further argument for the absence of dimension dependent 
powers of $2\pi$ in the constant
$c$, comes from an analysis of Feynman graphs in string theory and
quantum field theory.  The analog of the world sheet representation of
a string amplitude in field theory is the Schwinger parameter
representation of Feynman graphs.  This is derived from the momentum
space representation by doing all of the loop momentum integrals
for fixed values of the Schwinger parameters.  Each loop integral
contributes a factor of ${1\over {(4\pi)^{d\over 2}}}$ obtained from the 
explicit momentum space measure and the gaussian integral over momenta
(d is the spacetime dimension).  The origin of this factor is basically
the normalization of wave packets to one particle per unit volume, plus
the fact that particles must meet at a point in order to interact.
This factor multiplies the Schwinger parametric integrand.  By contrast,
in string theory we have only a single momentum integral multiplying
the integral over moduli space at any genus.  Presumably this is a
consequence of the fact that strings do not have to meet at a point in
order to interact.  Thus, {\it a priori} the expansion parameter
in string theory is $g^2$ rather than the coupling divided by a
dimension dependent power of $4\pi$.  

Of course, if one writes the
integral over moduli space in terms of sums over conformal field theory
states propagating in tubes (a string field theory like decomposition),
one can see a possible origin for multiple momentum integrals in string
theory.   The point of our argument is that, as a consequence of 
the more compact representation in terms of a partition function
integrated over multigenus moduli space, the existence of inverse powers
of $2\pi$ is much less obvious than in field theory (even in field
theory one could imagine that the parametric integrals gave compensating
powers of $2\pi$ in the numerator).  Further, the intuitive field
theoretic argument for the geometric factors is based on locality, which
is maximally violated in the \lq\lq fat'' part of moduli space which
gives rise to the large order behavior of string perturbation theory.
What we really need to settle this issue is a refinement of Shenker's
estimate of the large order behavior.

The bottom line of all of this speculation is a simple 
model of dilaton stabilization by an exponential superpotential and a
complicated nonperturbative kahler potential.  It is amusing that this
mechanism of stabilization of the dilaton can work, and give zero
cosmological constant, only if SUSY is broken.  This is a consequence of
a more general principle.  A nonzero superpotential (in the vacuum
state) is the order parameter for R-symmetries larger than the
$Z_2$ which simply reflects the supercoordinates.  With an exponential
superpotential, any nonzero value of the coupling thus implies broken
R-symmetry.  The standard supergravity formula for the potential
\eqn\pot{V = e^K(\vert D_i W\vert^2 - 3\vert W\vert^2)}
then implies that if R-symmetry is broken, SUSY must also 
be broken if the cosmological constant
vanishes.  In the weak coupling framework advocated here, SUSY breaking
is an automatic consequence of the dynamical principles of dilaton
stability and vanishing cosmological constant.

To my mind, this is the simplest of all possible models of SUSY breaking
in string theory.  Nonperturbative dynamics will generically lead to 
a nonvanishing superpotential (see below), which is guaranteed to be
dominantly a simple exponential in the weak coupling regime.  If the
Kahler potential can stabilize the dilaton with vanishing cosmological
constant in this regime, then SUSY is broken.  The price we must pay for
this attractive scenario is a certain impotence with regard to explicit
computations. Our assumptions imply that in the regime of interest we
cannot calculate the detailed form of the potential on moduli space.
In the cosmological context to which we now turn, this will have some
advantages (of the same nature of the advantages of the proverbial fig
leaf.).  Explicit perturbative superstring computations, combined with
exponential superpotentials, lead to potentials which are not compatible
with slow roll inflationary cosmology.

\newsec{Modular Cosmology}

The study of the cosmological effects of superstring moduli is necessary
because moduli typically lead to problems in cosmology.  It is
attractive because conventional slow roll inflation cries out for fields
with the properties of moduli.  The fundamental assumption of slow roll
inflation is that cosmological friction, a term of order $M_P^{-1}$
in the equations of motion, can compete with the force coming from
the gradient of the potential.  In conventional effective field theory,
in which field values of order the Planck scale are forbidden, potential
gradients of order $M_P^{-1}$ can only be achieved by unnatural fine tuning of
dimensionless parameters.  Even if one ignores the effective field
theory restriction on field space, as in theories of chaotic
inflation\ref\linde{A.D.Linde, {\it Phys. lett.}{\bf 129B},(1983),177.},
then one must fine tune what appear to be
dimensionless parameters in order to obtain density fluctuations of
small enough amplitude.  

By contrast, in string theory the natural range
of variation of the moduli is the Planck scale.  ${\nabla V \over V}$ is
always of order $M_P^{-1}$.  Furthermore, the potential vanishes to all
orders in perturbation theory, and is nonperturbatively bounded\bd by
something of order $e^{- {8\pi^2 \over {k g^2}}}$, where $k$ is an
integer which cannot be too large ($k$ is related to the size of the low
energy gauge group, whose rank is bounded in string theory).
If a nonperturbative potential is generated on moduli space then the
Hubble friction will naturally be of the same order of magnitude as the
restoring force and the amplitude of density fluctuations will naturally be
much less than one.  The order of magnitude estimate one obtains for the
number of e-foldings of inflation is $1$.  We know too little about the
lagrangian for the moduli to decide whether the phenomenological
constraint of more than $60$ e-foldings poses a fine tuning problem.

The notion that string theory moduli are inflatons
\ref\binet{P.Binetruy, M.K.Gaillard, {\it Phys.Lett.}{\bf
195B},(1986),382. } also
leads to a simple explanation of why we only see three large space
dimensions.  Moduli space is noncompact, but has finite
volume\ref\moorehorne{J.Horne, G.Moore, {\it Nucl.Phys.}{\bf B432},(1994),109.}
 .  For any finite value of the moduli, the
universe has finite volume and we should in principle seek a quantum
mechanical wave function for even the zero modes of the moduli fields.
It is only in the regions of moduli space corresponding to large volume
universes that we obtain approximate decoherence of states with
different values of the homogeneous modes.  

Let us try to sketch out
the quantum mechanics of the homogeneous modes by first neglecting both
the inhomogeneous modes and gravity.  This is the approximation of
quantum mechanics on moduli space.  For moduli an order of magnitude or
so larger than the string scale, we might argue that the effect of
inhomogeneous modes would be just to renormalize the metric and
potential on moduli space.  This renormalization should go to zero as we
approach the region of moduli space corresponding to large volume of 
more than three dimensions, because all low energy interactions are
infrared free.  
In these extreme regions of moduli space then, the lagrangian of the
moduli is well approximated by free motion in the classical metric.
This system has a normalizable eigenfunction which is a constant on
moduli space.  Since the volume of moduli space is finite, the
probability of being at large radius is predicted to fall like a power
of the radius.  Thus, any macroscopic universe is highly improbable.

This argument fails in four (or fewer) spacetime dimensions.  Well understood
effects in low energy nonperturbative field theory can lead to a
nonperturbative superpotential on moduli space even in the limit of
infinite three dimensional volume.  Once such a potential is generated,
the coupling of gravity, via the phenomenon of inflation, can invalidate
the above argument.  In dimension higher than four, the only scale for
the potential of scalar fields is the Planck scale.  Thus there is no
low energy effective field theory regime in which we can show that
inflation occurs.  Further, even if we could justify the use of
classical field theory, Planck scale inflation would lead to enormous
density fluctuations, and a highly inhomogeneous universe filled with
black holes.  The inflationary escape from the prediction of a small
universe is not available in high dimension string theory.

In order to prove that inflation occurred, one
would have to have more detailed information about the wave functional
of the inhomogeneous modes of the field.  One would want to show that
the energy
density was dominated by the potential energy of the constant mode over 
a region of space of size greater than the horizon volume corresponding
to that potential energy.  Thus we have not shown that inflation is
inevitable in four dimensions, only that it does not occur in more than
four, and that in its absence, string theory predicts the size of the
world to be very small.

Our basic contention then is that string theory predicts a very small
probability for the universe to be many orders of magnitude larger than
the string scale unless there are regions of moduli space in which
inflation occurs.  On the other hand, we will see below that the 
EFTS approximation probably breaks down if there are too many e-foldings
of inflation.  Furthermore, dimensional analysis (and some knowledge of
what happens in noncompact regions of moduli space) also suggests that
string theory cannot lead to too many e-foldings of slow roll inflation.
The order of magnitude estimate is $O(1)$.  

\subsec{Density Fluctuations and SUSY Breaking}

We have noted above that the string theorist's expectations for
nonperturbative modular physics lead to a plausible model of inflation.
For moduli we expect a potential of the form
\eqn\modpot{V = M_P^4 \sum e^{- c_i Re S}V_i ({\Phi})\equiv
M_i^4 V_i  .}
Where $S = {8\pi^2 \over g^2} + i\theta ; \quad 0\leq\theta\leq 2\pi $.
is the dilaton/axion field and $\Phi$ is a notation for all of the
dimensionless moduli fields including $ S$.  
The $V_i$ are supposed to be slowly
varying functions of $Re S$ in the weak coupling region.  In \bd it was
argued that truly stringy nonperturbative contributions were
characterized by $c_i = $even integer, while nonperturbative low energy
field theory can give fractional values of $c_i$.  Assuming that the
potential corresponding to the smallest $c_i \equiv c_1$ can lead to inflation,
the order of magnitude amplitude of density fluctuations is
\eqn\deltarho{{\delta\rho\over\rho} \sim ({M_1 \over M_P})^2 .}
Observation thus suggests $M_1 \sim 10^{-2.5} M_P$.

This is much larger than the highest allowed value for the square root of the
SUSY breaking order parameter $F$ .  If we assume that SUSY provides the
resolution of the weak interaction gauge hierarchy problem then
$\sqrt{F} < 10^{-8} M_P$.  The simplest resolution of this discrepancy
is to assume that the leading order potential $V_1$ has a supersymmetric
minimum.  In the inflationary cosmological context which we are
discussing, this has dramatic consequences.  The supergravity formula
for the potential \pot is generically negative for supersymmetric vacua.
Finding a supersymmetric stationary point requires us to solve $N$
complex equations for $N$ complex unknowns.  Requiring the
superpotential to vanish as well is an additional equation, which will
not generally be satisfied.

In the inflationary context, the resulting negative cosmological
constant is more than just a phenomenological embarrassment.  Inflation
has made the spatial curvature terms in Einstein's equations completely
negligible.  As a consequence, {\it there are no solutions of
the postinflationary cosmological equations in which scalar fields come
to rest at a minimum with negative cosmological constant}.  Instead, if
the system tries to relax to such a stationary point, the
universe recollapses on microscopic time scales, and inflation  
is not successfully completed.  Thus, the only supersymmetric minima 
into which the universe can settle after inflation, are nongeneric ones
where the superpotential vanishes.

String theory can provide us with such nongeneric superpotentials, but
as might be expected, they can only occur at special points in moduli
space.  First let us examine the simplest nonperturbative field theory
mechanism for generating a superpotential on moduli space, gaugino
condensation.  This leads to a superpotential of the form
\eqn\ggnosupot{W = e^{- CS} w({\cal M})}
Here ${\cal M}$ are the nondilatonic moduli, and the function $w$ is
determined by a one loop vacuum polarization calculation.  $C$ is
inversely proportional to the beta function of the hidden sector gauge group.

More generally, the low energy gauge theory which determines the leading
contribution to the superpotential on moduli space will contain massless matter
fields.  In many cases, the low energy contribution to the superpotential will 
favor infinitely large values of matter field VEVs.  Generically
however, there will be terms of dimension four or higher in the tree
level superpotential which will stabilize these VEVs at a finite value.
The result of integrating out the matter fields 
will be a superpotential for the moduli of the form
\eqn\gensupot{(e^{- C S} w({\cal M}))^p F({\cal M}) = \Lambda_{low}^p
F({\cal M})}
The second form of this equation is rewritten in terms of the
nonperturbative scale $\Lambda_{low}$
of the low energy gauge theory, which depends
implicitly on the moduli.  The power $p$ is
determined by the balance between the nonperturbative superpotential for
the matter fields, and the tree level term, and is always positive.  Finally, 
the function $F$ incorporates the moduli dependence of the coefficient
of the leading term in the tree level superpotential; the term which
stabilizes the massless modes of the matter fields..

A mechanism for obtaining a vanishing superpotential becomes apparent if
we imagine that at some point ${\cal M}_0$ 
in moduli space a number of other charged
chiral superfields become massless.  The effective value of
$\Lambda_{low}$ is then determined by renormalization group
matching\ref\seibint{K.Intriligator, N.Seiberg, {\it Lectures on
Supersymmetric Gauge Theories and Electric-Magnetic Duality},
RU-95-48, hep-th/9509066. } to a theory above the scale of the masses of
these fields but below the string scale.  If there are enough new
massless states at ${\cal M}_0$, then $\Lambda_{low}$ vanishes as the
masses go to zero.  If there is no compensating divergence in
the function $F$ then the superpotential will vanish at ${\cal M}_0$.
If it vanishes rapidly enough, ${\cal M}_0$ will be a supersymmetric
minimum of the potential.

Even without going into details of specific models, we can make some
general remarks about this mechanism.  It probably cannot work for
$(2,2)$ compactifications, where general theorems restrict the number
of massless chiral multiplets charged under the low energy gauge group.
The extant examples of this mechanism \ref\bdtwo{T.Banks, M.Dine, 
 {\it Quantum Moduli Spaces of N = 1 Vacua}, hep-th/9508071, to be 
published in Phys. Rev. D}
occur in vacua obtained
from a tree level $(2,2)$ vacuum by shifting fields to cancel the
Fayet-Iliopoulos term of a tree level anomalous $U(1)$ gauge
symmetry\foot{In passing we note that these vacua have been argued to be
exact quantum eigenstates of string theory.  They are supersymmetric,
have zero vacuum energy and are part of a continuous moduli space.
It is a challenge for string theorists to find an argument that explains
why the world is not sitting in such a state.}.

Returning to our model of inflation, we note that since the 
inflationary superpotential depends only exponentially
on the dilaton, the vacuum energy at the minimum 
is zero independent of the dilaton
field.  This is an example of the lesson we learned in the first
section.
The dilaton cannot be frozen by a SUSY preserving superpotential with
vanishing cosmological constant.

It is reasonable to assume, though by no means guaranteed, that the
nondilatonic moduli all obtain mass from the potential $V_1$.  This
means that their masses are of order $10^{-5} M_P$ and their nominal
reheat temperature of order $10^{- 7.5} M_P$.  This is high enough
to give us reason to worry about generating enough gravitinos at reheating to
cause problems with nucleosynthesis.  Perhaps we should not take this
too seriously.  We have been careless about numerical factors, and have
ignored the difference between the Planck scale and the string scale
(whose ratio might come into the estimate of the reheat temperature
raised to a rather large power).  What is clear is that if the
nondilatonic moduli are frozen by the potential which generates
inflation, they do not pose a reheating problem for the universe.
Temperatures high enough for baryogenesis and nucleosynthesis are
achieved by thermalization of the decay products of these moduli.

\subsec{Dilatonic Dilemmas}

We now come to the cosmological problems associated with the dilaton.
The first of these is the cosmological version of the Dine-Seiberg problem.
Brustein and Steinhardt\ref\bs{R.Brustein, P.J.Steinhardt, {\it
Phys.Lett.}{\bf B302},(1993),196, hep-th/9212049.} 
argued that even if a potential for the dilaton
was generated by SUSY breaking (and we have argued above that there is
no alternative in the weak coupling region), the energy in the dilaton
zero mode at early times would generally send it flying over this
tiny barrier into the extreme weak coupling region.  While the conclusion
is far from obvious even in a general model\foot{See the appendix of
\ref\bbs{T.Banks, M.Berkooz, P.J.Steinhardt, {\it op. cit.}}}, 
our present scenario provides a rather neat and definitive resolution
of the problem.  The dilaton will be stabilized by the second term
in \modpot , which, as in section 1, is also the term responsible for
SUSY breaking.  This potential will have a minimum at $S_0$. 
However, during inflation, the first term in \modpot
 will give a much larger effective potential for the dilaton
as long as the moduli ${\cal M}$ are far from their minimum.  Since the
potential is an exponentially varying function of the dilaton, the
dilaton will {\it not} be one of the inflaton fields.  In other words,
during inflation it will quickly be drawn into a local minimum $S({\cal
M})$ which depends on those of the other moduli which are undergoing
frictionally dominated motion.  If the end of this ``dilaton groove'',
$S({\cal M}_0)$ lies on the strong coupling side of the minimum $S_0$,
then the Brustein-Steinhardt
problem is solved.  The dilaton is gently deposited on the strong
coupling side of its true minimum at the end of inflation.  Its post
inflationary energy is of the order of the barrier height between the
minimum and extreme weak coupling.  

We do not have the tools at present to calculate the dilaton potentials
and so we cannot say that string theory definitely avoids the 
Brustein-Steinhardt problem.
However, we see that some simple qualitative properties of the potential
are sufficient.  We cannot with equal ease solve the problem of dilaton
domination of the energy density of the universe\ref\PPP{T.Banks,
D.Kaplan, A.Nelson, {\it op. cit.}; B. deCarlos, J.A. Casas, 
F.Quevedo, E.Roulet,\break {\it Phys.Lett.}{\bf B318},(1993),447,
hep-ph/9308325.}.  The dilaton
mass, since it is associated with hidden sector SUSY breaking, is of
order ${F\over M_P} \sim 10^{-16} M_P$.  Its reheat temperature is thus
of order $.01 MeV$ and one is hard put to understand the success of
conventional nucleosynthesis calculations.  The most promising solution
to this problem has been suggested by Lyth and Stewart\ref\ls{ D.Lyth,
E.Stewart, {\it Phys.Rev.Lett.}{\bf 75}, (1995),201,hep-ph/9502417; 
{\it Thermal Inflation and the Moduli Problem}, hep-ph/9510204.}.  
They argued that in many simple models, there would be a period of
$o(10)$ e-foldings of {\it thermal inflation}, with Hubble constant of
order the weak scale.  Randall and Thomas\ref\rt{L.Randall, S.Thomas, {\it Nucl.Phys.}{\bf
B449}, (1995), hep-ph/9407248 .} had argued that
such an era of intermediate scale inflation could redshift away the
dilaton (and gravitino) energy densities, without affecting the
observable properties of the Cosmic Microwave Background Anisotropy.
The advantage of the Lyth-Stewart idea is that during {\it thermal
inflation} the dilaton is attracted to a point quite close to its true
minimum, so that there is no post-inflation Polonyi problem.
Lyth and Stewart also suggest that their scheme may be able to
accomodate an adequate amount of baryogenesis.

Another, more speculative idea, is the assumption that the order of
magnitude estimate of the dilaton reheat temperature is simply wrong by
two orders of magnitude.  Then the dilaton could decay before
nucleosynthesis.  If there are renormalizable baryon number violating
operators in the supersymmetric standard model, dilaton decay could also
be the agent of baryogenesis.  In such a scenario, the small size of the
baryon to photon ratio is primarily explained by the large ratio between
the dilaton mass and its reheat temperature.  Such a model might be
compatible with conventional cosmology, but the required violation
of R-parity might deprive us of a natural candidate for dark matter.

Let us also note an issue that has not been addressed by any of the
attempts to avoid dilaton domination of the universe\bbs .  In all extant
models, there is a period, stretching roughly from the time the scale of
the energy density is $10^{11}$ GeV, until the weak scale, during
which
the dilaton or some other coherent scalar
does dominate the energy density.  Energy density
fluctuations on scales smaller than the initial horizon, grow by a
factor of almost $10^{11}$ during this era.  Since they initially had
amplitude of order $10^{-5}$, they go nonlinear long before the weak
scale is reached and the mechanisms for getting rid of the dilaton go
into effect.  The consequences of these nonlinear gravitational
phenomena have not been worked out, and until they are, any picture of
this era will be accompanied by a large question mark.

Finally, let me note that all resolutions of the problem of dilaton
domination work only for hidden sector SUSY breaking.  In a theory with
low energy SUSY breaking only, the dilaton will be extraordinarily
light.  Indeed, for reasonable values of the low energy SUSY breaking
scale ($ 1 - 100$ TeV), it is light enough to be on the edge of
detection by Cavendish experiments.
In such a model, the dilaton
is stable, and its energy density cannot be inflated away until long
after nucleosynthesis.  Thus, all string theory models with low energy
SUSY breaking have a serious unsolved cosmological problem.  

\newsec{The Cosmological Constant}

A major lacuna in any discussion of cosmology is of course the question
of why the cosmological constant is so small.  We have provided an
explanation above of why stationary points of the effective potential
with negative cosmological constant (and in particular, typical
supersymmetric minima) cannot be accessed after inflation.  There is no
corresponding mechanism for vacua with positive cosmological constant.

I believe that the answer to this question goes beyond the low energy
effective field theory description of string theory that we have used in
this lecture.  The main feature which distinguishes DeSitter space from
flat space is the existence of a horizon.  The conventional low energy
field theory picture of DeSitter space consists of an ever growing
number of causally disconnected regions, each with its own independent
set of field theoretic degrees of freedom.  Recent work of Lowe {\it et.
al.}\ref\noncommute{D.Lowe, J.Polchinski, L.Susskind, L.Thorlacius,
{\it Black Hole Complementarity Versus Locality}, 
NSF-ITP-95-47, hep-th/9506138.}
 suggests that such a picture is a very bad approximation to string
theory even when the cosmological constant is small and spacetime is
locally flat.  The discussion of \noncommute was oriented towards black
holes, but their actual calculations referred primarily to the low
energy effective description of a certain set of (nice slice)
coordinate systems in
flat spacetime.  These are a model of smooth coordinates which cross the
horizon and interpolate between the external Schwarzchild coordinates 
and some good set of coordinates in the interior of a large black hole.
A similar picture is sure to be approximately valid in DeSitter space.
The major technical problem in trying to apply these ideas to DeSitter
space, is that DeSitter space is not a solution of
the classical equations of string theory.

The message of \noncommute is that the low energy (with respect to
nice slice coordinates) effective description of string theory is not a
local field theory.  There are low energy states which are highly
nonlocal in nice slice coordinates.  They correspond to long strings
stretched from one side of the horizon to another.  Furthermore, there
is no subset of local measurements which shows independence
of the degrees of freedom on one side of the horizon from those on the other.  
Susskind argues that this is evidence for the holographic\ref\holo{G. 
't Hooft, {\it Dimensional Reduction in Quantum Gravity}, in Salamfest 1993,
gr-qc/9310026;
L.Susskind,{J.Math.Phys.}{\bf 36},(1995),6377.}
nature of string theory.  The theory has no bulk degrees of freedom.

If one accepts these arguments, there are several clear implications.
First of all, if an effective field theory calculation predicts a
positive cosmological constant in string theory, then it is not self
consistent.  The effective field theory is not a valid description of
low energy string theory on scales much larger than the DeSitter horizon.
(It also breaks down very near the horizon in static coordinates
which cover only a single horizon volume).  Secondly, if there are
really no bulk degrees of freedom, no effective field theory calculation
of the cosmological constant is correct in string theory.

These ideas are very crude and we surely do not yet have an accurate
idea of what they mean.  I would like nonetheless to give the first
outline of how one might calculate the cosmological constant in 
holographic string theory.  I will begin by assuming that string theory
describes our present universe as a roughly spherical object of radius
$R$ (by which I mean only that its volume is approximately the three
halves power of its area).  
Furthermore, the true degrees of freedom of the system are assumed
to be distributed on a two dimensional submanifold of area $R^2$. This
is the transverse submanifold of some light cone frame.
 Part
of the magic of the theory, which we are very far from understanding, is
that we can choose to locate the degrees of freedom on any submanifold
with the same area.  The density of degrees of freedom of the system
on the submanifold is assumed to obey the Bekenstein rule: one
degree of freedom per Planck area.  

On the other hand, we know that the Universe can be described by local
field theory, to some degree of approximation.  Indeed, one might argue
that since we have tested local field theory up to energies of order
{\it e.g.} $1$ TeV, we know that the universe must contain at least
$(1\  TeV)^3 R_{hor}^3$ degrees of freedom ($R_{hor}$ is the horizon
scale).  This is incompatible with the holographic principle, but the
argument is not really valid because the holographic degrees of freedom
are related to local measurements in an extremely nonlocal way.  We can
only falsify the holographic idea by simultaneous measurements
throughout the universe.  If we tried to build a network of accelerators
to make this test at the TeV scale, 
we would instead form a collection of black holes
which swallowed up the apparatus.

I believe however that it is reasonable to claim that the degrees of
freedom represented by the microwave background radiation today are
independent.  There is no problem of imagining a network of experiments
to probe them, without forming black holes.  We may ask, for what radius
$R_U$ of the universe are the microwave oscillators equal in number to a
Planck density of degrees of freedom spread over an area of order
$R_U^2$.  The answer is $R_U \sim 10^{35} R_{hor}$.\foot{The argument
in the previous two paragraphs is an interpretation of remarks
of L. Susskind.}

We can now try to calculate the cosmological constant.
The total light cone energy of the
system (which will be the cosmological term) is given by summing up the
zero point fluctuations of these transverse degrees of freedom.  
We will also assume supersymmetry, broken at a scale $\sqrt{F}$.  The
contributions of bosons and fermions with energies and transverse
momenta
greater that $\sqrt{F}$ will cancel.  The total energy is then
\eqn\etot{E = F^{3\over 2} R_U^2 .}
with everything measured in Planck units.
But we claim that for this value of $R_U$, the system can be
reinterpreted as an effective local field theory spread over the volume.
The cosmological term in the light cone energy is thus reinterpreted as
the volume integral of a cosmological constant of order
\eqn\cosmocon{\Lambda \sim {F^{3/2} \over R_U}}
If $\sqrt{F} \sim 10^{- 8} M_P$, as we would expect in hidden sector
models of SUSY breaking, then this gives a cosmological constant of
order the current bound if $R_U\sim 10^{96} M_P^{-1}$.  This is about
$10^{35}$ times as large as the radius of our horizon volume, precisely
the radius fixed by consistency between the microwave background entropy
and a holographic picture.
TeV scale SUSY breaking gives a cosmological constant $10^{-24}$ times
smaller than the current bound, for the same radius $R_U$.

One amusing feature of this calculation is that it leads us to expect a
nonzero value for the cosmological constant.  This is of course favored
by some fashionable cosmological models. In the above calculation, one obtains
a phenomenologically interesting value of the cosmological constant only
for hidden sector SUSY breaking.

There are many features of the preceding argument which are obscure.
In particular, one may ask what is special about the present moment
in cosmological history.  Does our calculation imply that the
cosmological constant and/or the number of degrees of freedom in the
universe changes with time?  I do not have any clear idea of how to
answer either these or a host of other questions.

\newsec{Patterns of SUSY Breaking}

The model of SUSY breaking advocated in Section 1 implies that the SUSY
breaking order parameter is the dilaton F term.  In previous work on
perturbative string theory such an assumption was shown to lead to a
highly constrained pattern of SUSY breaking.  Within the framework of
\bd however, most of this predictive power is lost.  Consider for
example the squark masses.  The Kahler potential will have terms of the
form $Q^{\dagger}_i (M_Q^2 )^{ij} Q_j$ where $i$ and $j$ are generation
indices.  The matrix $M_Q^2$ will be dilaton dependent.  Within the
framework of \bd one can make no general statements about it in the
coupling regime of interest.  In particular, there are no apriori
relations between the matrix evaluated at the dilaton VEV, and its mixed
second derivative (wrt the dilaton) 
evaluated at the same point.  These two matrices
determine the squark wave function renormalization and the doublet
squark masses respectively.  Thus without imposing discrete symmetries,
there is no reason for the squark mass matrix to be naturally
proportional to the unit matrix.  Similar remarks are valid for up and
down squark mass matrices.  

Next consider the $\mu$ and $B$ parameters of the supersymmetric
standard model.  The absence of the $\mu$ term in the tree level
superpotential might be the consequence of a stringy symmetry, or of an
R symmetry under which the operator $H_u H_d$ carries charge zero.
In either case there is nothing to forbid a term $ G H_u H_d + h.c.$, where
$G$ is a function of the dilaton field, in the Kahler potential.
The dilaton F term will then generate both a $\mu$ and a $B$ term of the
right order of magnitude.  They will be related to the first and second
derivatives of $G$ at the dilaton VEV and there is no particular
connection between them.   One can also obtain a contribution to
the $\mu$ term from a tree level coupling of $H_u H_d$ in the gauge
kinetic function of the hidden gauge group\ref\taylor{
I.Antoniadis, E.Gava, K.S.Narain, T.R.Taylor, {\it Nucl.Phys.}{\bf
B432},(1994),187, hep-th/ \break 9405024.}.  

If the symmetry which forbids $H_u H_d$ in the tree level superpotential
is not an R-symmetry, or is an R-symmetry under which this operator
carries nonzero charge, then the mechanism described above does not
work.  The $\mu$ and $B$ terms could only be obtained by coupling to a
standard model singlet, which carries the relevant quantum number, 
in the tree level potential.  

The only one of the weak coupling predictions of dilaton dominated SUSY
breaking which survives in the scheme of \bd is the prediction of
universal gaugino masses.  The gauge kinetic functions are indeed
unrenormalized apart from extremely small terms, and their dilaton
dependence is determined at tree level.

It is also appropriate here to correct some statements that were made in
\bd about the mass of the model independent axion.  There it was claimed
that the axion mass could naturally be very small.  In fact, there is no
principle known to me which could prevent the occurrence of a term in
the Kahler potential of the form $e^{- c \sqrt{Re S}} f(S,S^* )$.
According to the rules of \bd , this term must be considered of order
one.  In general it will give the axion a mass of order the gravitino
mass.  No discrete symmetry can alter this conclusion.  A gauged
anomalous $U(1)$ forces the Kahler potential to be a function of only
$Re S$, plus terms involving charged fields.  However, in all known
examples, the anomalous $U(1)$ is spontaneously broken to a discrete 
subgroup by the VEV of a
charged field only slightly smaller than the string scale\foot{If this
does not occur, the whole dilaton supermultiplet is Higgsed and
disappears from the spectrum at a scale much higher than that of SUSY
breaking. None of our discussion would be valid in such a hypothetical
situation.} .  Thus, the axion mass can at most be suppressed by a few
powers of the charged field VEV over the string scale.  It will never be
light enough to serve as a QCD axion, or as dark matter. 

\newsec{Conclusions}

Let us summarize our conclusions as a series of tentative answers to popular
questions:

\noindent
1.{\bf Why Is SUSY Broken?} - The answer is essentially cosmological.
Inflation is necessary to obtain a large universe.  Postinflationary
universes do not settle into vacua with negative cosmological constant.
If R-symmetry is broken (and this happens in a very large class of
strongly coupled
supersymmetric gauge models even when SUSY is preserved) then
supersymmetric vacua have negative cosmological constant.  The criterion
for zero cosmological constant is instead {\it that SUSY be broken, and that
there be a certain relation between the SUSY breaking scale and the
R-symmetry breaking scale}.  In hidden sector models of SUSY breaking,
this relation is automatically satisfied in order of magnitude.  

In models of low energy SUSY breaking, the vanishing of the cosmological
constant is more of a puzzle.  It is traditional to resolve it by simply
adding a constant to the superpotential.  I do not think this is a
satisfactory answer.  A constant in the superpotential breaks
R-symmetry, and we should understand the scales at which symmetries are
broken in a dynamical manner, rather than putting them in by hand. 
In \bbs a technically natural solution to this problem was proposed.  
The negative contribution to the potential from a strongly coupled
supersymmetry preserving sector could have the same order of magnitude
as the positive contribution from low energy SUSY breaking if the beta
function of the SUSY preserving gauge group was $1.5$ times as large as
that of the SUSY breaking one.  Although this is technically natural, it
seems somewhat contrived.  The \lq\lq natural in order of magnitude''
cancellation of the cosmological constant might be advanced as another argument
which favors hidden sector SUSY breaking (recall that in models with low
energy SUSY breaking, the dilaton is extremely light and dominates the
energy density of the universe in an unacceptable manner).

\noindent
2.{\bf How is the Dilaton Stabilized?} -  In string
theory, any nonperturbative dynamics will produce a superpotential for
the dilaton which is a sum of exponentials in the weak coupling region.
The hierarchy of scales in nature is a consequence of these
exponentials.  In particular, we emphasize the hierarchy which is
necessary to successful inflation: if the energy density during a
large number of final e-foldings of inflation is not substantially
smaller than the Planck scale, then large energy density fluctuations
cause all of the matter in the universe to recollapse into a set of
black holes.  Thus, from a cosmological point of view, the dilaton must
be stabilized in the weak coupling regime.  Apart from cancellation
between different exponentially small effects (which we eschew for
religious reasons) there is one hypothetical mechanism for stabilizing
the dilaton at weak coupling.  If it is true that the perturbative
estimate of the Kahler potential is wrong in the region where low energy
field theory is weakly coupled, then it is also possible that the
dilaton potential has a stable weak coupling minimum with zero
cosmological constant even for a purely exponential superpotential.
As above, this is only possible if SUSY is broken.

\noindent
3.{\bf Why Do We Live in Four Dimensions?} - A partial answer is given to
this question by observing that in a finite universe the LEFTS is {\it
quantum mechanics on moduli space}.  In the region where more than four
spacetime dimensions are large, it is plausible that this is a good
approximation, since the correction to the action for homogeneous modes
from integrating out other modes,
including gravity, is under control in the infrared.  
Since the volume of moduli space is finite, the theory
predicts a probability for large dimensions which falls like a large
power of the size of the universe in string units.  In four dimensions,
this argument may be incorrect.  If potentials are generated on moduli
space by low energy dynamics, then we {\it may} have inflation.  Thus,
once the universe was a few orders of magnitude above the string scale,
inflation would blow it up to very large size.  We conclude that, {\it the most
probable large universes in string theory have four (or perhaps fewer)
dimensions}.  Some arguments have been given\bs that two dimensions are
also improbable so the problem is reduced to understanding why we do not
live in three dimensions.  

Note that it has not yet been argued that
inflation is highly probable.  Indeed, the {\it improbability} of
inflation in string theory might be part of the principle which
selects the string vacuum state.  That is, the reason that a particular
vacuum is chosen may be because it is the basin of attraction for one of
the rare(?) places in moduli space where inflation occurs.

\noindent
4.{\bf How is the String Vacuum Chosen?} - Here the answer given in \bbs is
less satisfactory and depends on quantitative phenomenological input
rather than the qualitative requirement that the universe be much larger
than microscopic size (though note the speculation at the end of the
previous answer).  Data on the microwave background radiation
suggest a potential energy density just a few orders of magnitude below
the Planck scale, much larger than that allowed by SUSY breaking.  Our
most successful explanation of this discrepancy involves a
superpotential generated by SUSY preserving strong dynamics at the high
scale. This has the added advantage that most of the string moduli (but
not the dilaton) will
become massive at a high enough scale to avoid problems with reheating.
The requirement that the low energy theory have such a SUSY preserving 
sector becomes a vacuum selection principle because of
the observation that (as in 1 above)
supersymmetry preserving dynamics can only lead to successful inflation 
if the superpotential satisfies a nongeneric condition. It must vanish
at the minimum of the potential.  In string
theory this condition is quite restrictive and can only be satisfied
at points in moduli space where chiral multiplets charged under the
hidden sector gauge group become massless (without enlarging the group).
Very few points satisfying this criterion are known (and as yet, none of
them have anything resembling satisfactory phenomenology). It is not
implausible to suggest that they are isolated points or lie in very
low dimension subspaces of moduli space.  This means that most of the
moduli are massive at these stationary points.  

Clearly this is at best the beginning of an answer to the question of
vacuum selection.  The vacuum selection principle we have invoked so far
depends on fitting a quantitative aspect of the known universe, the
amplitude of microwave background fluctuations. 
We would like to be able to show that all vacuum
states other than the one which fits the detailed phenomenology of our
world lead to clearcut cosmological disasters.  Thus one would like to
put only the barest phenomenological input into an explanation of vacuum
selection.  It may be that in the end we will have to accept a more
equivocal answer to this question.  Quantum string theory may predict a
reasonable probability for many different kinds of large universe.  In
this case fundamental physics would inherit part of the contingent,
historical character of evolutionary biology.  Certain features of the
world would just be lucky accidents.  One might attempt to invoke
some sort of anthropic principle to distinguish our world, but I despair
of ever being able to understand the physics of one of the \lq\lq
alternate universes'' well enough to prove that no sort of intelligent
life could ever evolve.  In the context of string theory as we
understand it today, the best hope for a clearcut vacuum selection
principle seems to me to be the problem of the cosmological constant.

\noindent
5.{\bf Why is the Cosmological Constant So Small?} - For negative values
of the
cosmological constant, we have given an answer to this question, but for
positive values I believe that the
answer goes beyond the effective field theory
approximation to string theory.  Lowe {\it et. al.} have shown that in
the presence of horizons, the effective low energy theory of string
theory is nonlocal, and contains states of long strings stretched along
the horizon.  This is true even when the spacetime curvature is small.
Indeed it is true in flat space in appropriate coordinate systems (there
is no contradiction with general covariance here if we believe that
string theory is essentially nonlocal).  {\it We conclude that
that if the low energy field theory approximation to string theory
predicts a positive cosmological constant, then it is not
self-consistent.}  Truly stringy dynamics will have to be invoked to
understand the dynamics of the universe on scales much larger than the
DeSitter horizon.  Note that this does not contradict the field
theoretic treatment of a sufficiently short period of inflation.
For a given value of the Hubble parameter, the field theoretic treatment
will be valid up to some maximum number of e-foldings. At present
I do not understand how to calculate this maximum number.
Thus, we may claim that in string theory we are completely ignorant of
the fate of vacuum states that are predicted to have positive
cosmological constant by low energy field theory approximations.

Ignorance is bliss.  The observation of Lowe {\it et. al.} allows us to
{\it imagine} that string theory will predict disaster for any vacuum
state which, according to low energy field theory, is a DeSitter space.
Disaster might mean recollapse, or some more exotic phenomenon which
does not even have a local spacetime description. If this is the case,
then the fine tuning problem of the cosmological constant could become a
virtue.  Only vacuum states which had very small cosmological constant
would survive, and these would surely be few and far between. 
Note the further constraint that the cosmological constant vanish in the
weak coupling regime.  A strongly coupled vacuum with vanishing
cosmological constant would not have any obvious mechanism for producing
a hierarchy.  In particular, one would expect the scale of primordial
${\delta\rho\over \rho}$ in such a vacuum (presuming it gives rise to some
sort of inflation) to be $0(1)$.  This leads to black hole
formation and cosmological disaster.  The idea that the vacuum is
selected by requiring vanishing cosmological constant at weak coupling
seems sufficiently powerful to pick out a unique state.  Indeed, from
the point of view of low energy field theory it seems suprising that
there is any solution to this constraint.  

However, we must remember that the work of Lowe {\it et. al.} may be 
taken to support the
conjecture\ref\lenny{L.Susskind, {\it op. cit.}} 
that string theory is holographic.  If string theory is truly as
nonlocal as the holographic principle implies, then our field theoretic
understanding of a natural value for the cosmological constant is surely
suspect. 
 We presented
a wildly speculative calculation of the cosmological constant based on
these ideas.  With hidden sector SUSY breaking, it gives a value close
to the present observational bound.

\noindent
5.{\bf What is The Pattern of Low Energy SUSY Breaking?} - In our model
for SUSY breaking, almost everything of interest is bound up with the as
yet unknown Kahler potential of the low energy fields.  The only clear
cut prediction is that gaugino masses are universal.  Flavor problems
will have to be resolved by a horizontal symmetry.  We noted that both
the dilaton and axion have masses of order $1$ TeV, so that the
strong CP problem will have to be solved by a vanishing up quark mass.
The nonperturbative Kahler potential will also change the predictions
for coupling constant unification, in a manner which is at present
uncomputable.  

More definite phenomenological predictions await a
nonperturbative calculational scheme for string theory.  It is possible
that something could be done by resumming the perturbation expansion.
If we could, by other means, decide on the correct point in moduli space
to expand about, it would be worth the effort to try to compute several
terms in the expansion of the Kahler potential.  Furthermore one can
probably also get a more precise determination of the large order
behavior of the series than we have at present.  Then, some variation of
the Borel transform/conformal mapping techniques which have been
successful in the theory of critical phenomena might be applied.  This
program sounds like an enormous amount of work, but given a likely
candidate for the ground state it is quite concrete, and perhaps feasible.

We also noted that low energy SUSY breaking models suffer from an as yet
unresolved cosmological reheating problem.  This is the consequence of
the fact that in such models the dilaton is extremely light.  Indeed it
is probably light enough to be experimentally discovered in direct
terrestrial measurements of the forces between objects.  It would be
extremely interesting to improve the bounds on the range of coherent
forces that are consistent with such experiments.  

The dilaton reheating
problem is one of two
indications that string theory will in the end lead us to a hidden
sector mechanism for SUSY breaking.  The other is the order of magnitude
cancellation of the cosmological constant which is \lq\lq automatic'' in
hidden sector theories and requires some tinkering in theories of low
energy SUSY breaking.

To summarize, we have proposed cosmological answers to many of the
fundamental questions facing superstring theory.  The basic cosmological
constraint is the existence of a vacuum state of the theory with vanishing
cosmological constant, stable dilaton, and broken R-symmetry.  SUSY
breaking (probably via a hidden sector mechanism) is a consequence of
these principles.  A further constraint is that the vacuum lie in the
basin of attraction of a portion of moduli space where inflation occurs.
We have argued that very large numbers of e-foldings are not expected
in the LEFTS, and that the low energy theory inevitably loses validity
({\it i.e.} is not a good approximation to string theory) when it
predicts a large number of e-foldings.  We do not yet have a sufficient
quantitative understanding of the theory to determine whether the
phenomenologically necessary $60$ e-foldings is a large or small number
from this point of view.  

Indeed much of the detailed quantitative information about the theory is
obscured in our approach by the necessity of invoking poorly understood
nonperturbative phenomena to stabilize the dilaton in the weak coupling
regime. This is an inevitable consequence of the Dine-Seiberg argument.
Any truly predictive framework for string theory requires us to
formulate and solve the theory in a nonperturbative manner.
\vfill\eject
\centerline{\bf ACKNOWLEDGEMENTS}

I would like to thank, M. Berkooz, M.Dine, G.Moore, L.Randall, S.Shenker,
P.J.Steinhardt, L.Susskind, and S.Thomas for numerous discussions about
the subjects touched on in this lecture. 

\listrefs
\end